\documentclass[aps,showpacs,twocolumn,superscriptaddress,longbibliography,floatfix]{revtex4-2}
\usepackage[colorlinks,linkcolor=blue,citecolor=blue,anchorcolor=blue,urlcolor=blue]{hyperref}
\usepackage{enumitem}
\usepackage{bm,amssymb,amsmath}
\usepackage{dcolumn}
\usepackage{graphicx}
\usepackage{multirow}
\usepackage{graphicx}

\begin{document}

\title{Universality of percolation at dynamic pseudocritical point}

\date{\today}

\author{Qiyuan Shi}
\affiliation{Department of Modern Physics, University of Science and Technology of China, Hefei 230026, China}

\author{Shuo Wei}
\affiliation{Department of Modern Physics, University of Science and Technology of China, Hefei 230026, China}

\author{Youjin Deng}
\email{yjdeng@ustc.edu.cn}
\affiliation{Hefei National Research Center for Physical Sciences at the Microscale, University of Science and Technology of China, Hefei 230026, China}
\affiliation{Department of Modern Physics, University of Science and Technology of China, Hefei 230026, China}
\affiliation{Hefei National Laboratory, University of Science and Technology of China, Hefei 230088, China}

\author{Ming Li}
\email{lim@hfut.edu.cn}
\affiliation{School of Physics, Hefei University of Technology, Hefei 230009, China}

\begin{abstract}
    Universality, encompassing critical exponents, scaling functions, and dimensionless quantities, is fundamental to phase transition theory. In finite systems, universal behaviors are also expected to emerge at the pseudocritical point. Focusing on two-dimensional percolation, we show that the size distribution of the largest cluster asymptotically approaches to a Gumbel form in the subcritical phase, a Gaussian form in the supercritical phase, and transitions within the critical finite-size scaling window. Numerical results indicate that, at consistently defined pseudocritical points, this distribution exhibits a universal form across various lattices and percolation models (bond or site), within error bars, yet differs from the distribution at the critical point. The critical polynomial, universally zero for two-dimensional percolation at the critical point, becomes nonzero at pseudocritical points. Nevertheless, numerical evidence suggests that the critical polynomial, along with other dimensionless quantities such as wrapping probabilities and Binder cumulants, assumes fixed values at the pseudocritical point that are independent of the percolation type (bond or site) but vary with lattice structures. These findings imply that while strict universality breaks down at the pseudocritical point, certain extreme-value statistics and dimensionless quantities exhibit quasi-universality, revealing a subtle connection between scaling behaviors at critical and pseudocritical points.
\end{abstract}

\maketitle

\section{Introduction}
\label{sec-intro}
Percolation theory~\cite{stauffer1991} offers a fundamental framework for studying diverse systems, including materials science~\cite{kim2016revisit}, complex networks~\cite{LI20211}, and epidemiology~\cite{ziff2021percolation}, where geometric phase transitions occur. The percolation model operates on a simple rule: each site or bond is occupied with a given probability $p$. As $p$ increases and exceeds the percolation threshold $p_c$, the system transitions from being dominated by small, disconnected clusters to one where a giant component emerges, connecting large portions of the system.

As a system approaches the percolation threshold $p_c$, a series of critical behaviors emerge, with certain physical quantities either diverging or reaching finite values, often characterized by critical exponents. Importantly, these exponents do not depend on the details of the system, such as the lattice type, site or bond percolation, instead, they are determined by the spatial dimension, the range of interaction, and symmetry. Models that share the same set of critical exponents are said to belong to the same universality class. 
In two dimensions (2D), exact results can be derived using methods such as lattice duality/matching~\cite{10.1063/1.1704215}, Yang-Baxter integrability~\cite{PhysRevLett.18.692,BAXTER1972193}, and local conformal invariance~\cite{BELAVIN1984333,PhysRevLett.52.1575}. 
Critical exponents, such as the correlation-length exponent $\nu=4/3$ and the magnetic exponent $y_h = 98/41$ (also known as fractal dimension $d_f$), have been predicted by Coulomb gas theory~\cite{nienhuis1984critical}, conformal field theory~\cite{CARDY1986219}, and stochastic L\"oewner evolutions~\cite{CARDY200581}, and strictly proven on the triangular lattice~\cite{smirnov2001critical}.

For a finite system of side length $L$, the finite-size scaling theory hypothesizes that there exists a critical finite-size scaling window $\mathcal{O}(L^{-1/\nu})$, and, within the window, the finite-size behavior of physical quantities $Q$ would behave as
\begin{equation}
Q(t,L)=L^{y_Q}\tilde{Q}(tL^{1/\nu}),     \label{eq-fss}
\end{equation}
where $t=p-p_c$ is the distance to the criticality, $y_Q$ is the critical exponent of $Q$, and $\tilde{Q}(x)$ is a universal function. For finite systems, the $L$-independent pseudocritical point $p_L$ has the asymptotic behavior $p_L=p_c+aL^{-1/\nu}$, where $a$ is an $L$-independent parameter, yielding $(p-p_c)L^{1/\nu}=(p-p_L)L^{1/\nu}+\mathcal{O}(1)$. This means that defining $t=p-p_L$ only introduces an $L$-independent shift for the argument in the scaling function $\tilde{Q}(tL^{1/\nu})$, which has no influence on the finite-size scaling. Thus, it is generally assumed that the finite-size scaling ansatz Eq.~(\ref{eq-fss}) is applicable both around $p_c$ and $p_L$.

It is noted that the definition of $p_L$ is not unique. At various pseudocritical points $p_L$ and critical point $p_c$, representing different ensembles, it is generally believed that the same finite-size scaling behavior can be observed. However, different finite-size behaviors have also been reported for $p_c$ and $p_L$ of different definitions. For example, the finite-size scaling of explosive percolation at $p_c$ is shown to be abnormal~\cite{Grassberger2011,Souza2015}, which was initially misidentified as a discontinuous phase transition~\cite{Achlioptas2009,Friedman2009,Ziff2009,Souza2010}. Recently, it is found that explosive percolation obeys the standard finite-size scaling theory around the dynamic pseudocritical point~\cite{PhysRevLett.130.147101,Li2024}, indicating the non-equivalence of ensembles. Another example is the high-dimensional percolation model, where the finite-size scaling is dependent on the boundary conditions~\cite{PhysRevE.110.044140,Kenna2017}. For free boundary conditions in $d$ dimensions, the fractal dimension at $p_L$ is $d_f = 2d/3$, in agreement with that under periodic boundary conditions~\cite{rsa.20051,ChristianRandom,heydenreich2007random}, while $d_f=4$ at $p_c$~\cite{AIZENMAN1997551,crossoverfinitesizescalingtheory}. Besides, a study also suggests that by defining a dynamic pseudocritical point, the observables, such as the size of the largest cluster, follow the extreme-value Gumbel distribution across different models~\cite{fan2020universal}.

In this work, we demonstrate that, even in 2D, where finite-size scaling behaviors at pseudocritical and critical points are generally considered identical, there may still be differences in their universality. Through extensive simulations, we observe that the probability distribution of the largest-cluster size asymptotically approaches a Gumbel distribution in the subcritical phase, a Gaussian distribution in the supercritical phase, and varies based on the sample position within the critical finite-size scaling window. At a pseudocritical point, this distribution retains a universal form across different lattices but differs from the critical point. Moreover, the critical values of certain dimensionless quantities, such as wrapping probabilities~\cite{langlands1992,pinson1994critical,ZIFF199917}, and Binder cumulants~\cite{PhysRevLett.47.693, binder1981finite}, still take finite constant values at pseudocritical points. However, their magnitudes are significantly different from those at the critical point. Our data suggest that these dimensionless quantities are identical for bond and site percolation, but different for various lattices. All these suggest that although universality at criticality may break at the pseudocritical point, some extreme-value statistics and dimensionless quantities can exhibit a form of quasi-universality, indicating a nuanced relationship between scaling behaviors at critical and pseudocritical points.

The remainder of this paper is organized as follows. In Sec.~\ref{sec-simulation}, the simulation details and sampled quantities are described. Section~\ref{sec-results} presents our numerical results. Finally, a brief summary and discussion are given in Sec.~\ref{sec-diss}.

\section{Simulation}
\label{sec-simulation}
\subsection{Models}
We study percolation on 2D lattices with the periodic boundary conditions, including
\begin{enumerate}[label=(\alph*)]
    \item Bond percolation (BP) on square and triangular lattices;
    \item Site percolation (SP) on square and triangular lattices;
    \item Equivalent-neighbor bond percolation (EBP)~\cite{PhysRevE.98.062101} on square lattices.
\end{enumerate}

\begin{figure}
    \centering
\includegraphics[width=0.48\linewidth]{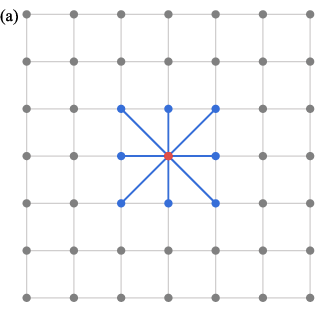}
\includegraphics[width=0.48\linewidth]{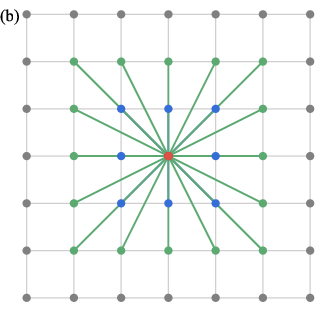} 
\caption{ Illustration of the EBP on the square lattice with different interaction range $r$. (a) 
For $r=1$, each site $i$ (e.g., the red site) connects to all its neighbors $j$ within the range $|x_i-x_j| \le r$ and $|y_i- y_j| \le r$ 
(indicated by blue), where $(x, y)$ is the site coordinates. Gray sites are outside the interaction range $r=1$ and cannot form bonds with 
the red site. (b) For $r=2$, in addition to the blue sites, more bonds (indicated by green) are formed with the red site. Under periodic 
boundary conditions, each site has $z=(2r+1)^2-1$ neighbors, and the total number of bonds in the system is $zL^2/2$. Each bond is occupied 
with an equal probability $p$, and unoccupied with probability $1-p$.
}  \label{fig1}
\end{figure}

For BP, each bond are occupied with a probability $p$, and the percolation threshold is 
$p_c=0.5$ and $p_c=2\sin(\pi/18)$ for square and triangular lattices, 
respectively~\cite{10.1063/1.1704215}. For SP, the occupied probability $p$ refers to site, and 
$p_c=0.59274621(13)$~\cite{PhysRevLett.85.4104} for square lattices, 
$p_c=0.5$~\cite{PhysRevE.60.6496} for triangular lattices. EBP takes the same rule as BP, 
but bonds can be inserted between sites within a distance of $2r\times 2r$, where $r$ is an adjustable 
parameter. Specifically, a site $i$ with coordinates $(x_i, y_i)$ on square lattice has a probability 
$p$ to couple to each site $j$ satisfying $|x_i-x_j| \le r$ and $|y_i- y_j| \le r$,  
as shown in Fig.~\ref{fig1}. For finite $r$, 
EBP increases coordination number without changing the spatial dimension, thus it also belongs to 2D 
percolation universality~\cite{PhysRevE.98.062101,deng2019medium}. Without loss of generality, we set 
$r=5$ so that a site has $120$ neighbors.

\subsection{Ensembles}

A conventional method for studying the finite-size scaling of percolation operates as follows: all occupied bonds or sites are introduced into the system simultaneously, with the occupation probability $p$ serving as the control parameter. This generates a configuration of occupied bonds or sites. By setting $p=p_c$, critical observables can be sampled and averaged over different realizations. This approach is referred to as the conventional ensemble.

In contrast, the event-based ensemble represents the percolation process as a dynamic bond or site insertion process~\cite{PhysRevLett.130.147101,Li2024,fan2020universal}. Initially, all bonds or sites are in an unoccupied state. Then, at each time step $T$, a bond or site is randomly selected and changed to the occupied state. By dynamically recording the size of the largest cluster, $\mathcal{C}_1(T)$, a special time step $T_{\text{max}}$ can be identified, where the gap $\Delta=\mathcal{C}_1(T)-\mathcal{C}_1(T-1)$ reaches its maximum value. Let $E$ denotes the total number of bonds and $V$ denotes the total number of sites. The time step $T_{\text{max}}$ corresponds to a pseudocritical point, $p_L = T_{\text{max}}/E$ or $p_L = T_{\text{max}}/V$, which fluctuates from realizations to realizations. Using the recorded sequence of occupied bonds or sites, the configuration at $p_L$ can be reconstructed, allowing any quantities of interest to be sampled. The ensemble average is then obtained as the mean of the observables measured at this dynamic pseudocritical point $p_L$.

\subsection{Sample quantities} 

For each realization of a percolation model, the following observables are sampled at both $p_c$ and $p_L$,
\begin{enumerate}[label=(\alph*)]
    \item The size of the $i$-th largest cluster $\mathcal{C}_i$. 
    \item The second moment of cluster-size distribution $\mathcal{S}_2 = \sum_i \mathcal{C}_i^2$ and the fourth moment $\mathcal{S}_4 = \sum_i \mathcal{C}_i^4$, where  the sum $\sum_i$ runs over all clusters.
    \item The indicators $\mathcal{R}_2$ and $\mathcal{R}_0$ for the existing of wrapping clusters. $\mathcal{R}_2 = 1$ when having a cluster wrapping along both directions, otherwise $\mathcal{R}_2 = 0$. Similarly, $\mathcal{R}_0=1$ when 
    having no wrapping cluster, otherwise $\mathcal{R}_0=0$.
\end{enumerate}

From these observables, we can calculate the following quantities,
\begin{enumerate}[label=(\alph*)]
    \item The mean size of the $i$-th largest cluster $C_i = \langle\mathcal{C}_i\rangle$. 
    \item The probability distribution $f_{C_i}(x)$ for the size of the $i$-th largest cluster with $i = 1,2,3$.
    \item  The second order Binder cumulant $B_2 = \langle \mathcal{C}_1 \rangle^2 / \langle \mathcal{C}_1 ^2\rangle$. 
    \item  The fourth order Binder cumulant $B_4 = \langle \mathcal{S}_2 \rangle^2 /( 3\langle \mathcal{S}_2^2
    \rangle - 2 \langle \mathcal{S}_4 \rangle)$. 
    \item  The wrapping probabilities $R_2 = \langle \mathcal{R}_2 \rangle $ and 
    $R_0 = \langle \mathcal{R}_0 \rangle $.
    \item The critical polynomial $P_B = \langle \mathcal{R}_2 - \mathcal{R}_0 \rangle $. 
\end{enumerate}
Here, $\langle \cdot \rangle$ denotes the ensemble average.

\section{Results}
\label{sec-results}
\subsection{The size distribution of the largest cluster}

The percolation transition is often characterized by the size of the largest cluster $C_1$, which scales as $C_1 \sim \ln V$ below $p_c$, $C_1\sim V$ above $p_c$, and $C_1\sim L^{d_f}$ at criticality. For a occupied probability $p$, the largest-cluster size varies across different realizations. To investigate the universal behavior of $C_1$, we first examine the size distribution $f_{C_1}(x)$ of the largest cluster. 

\begin{figure*}
\centering
\includegraphics[width=\linewidth]{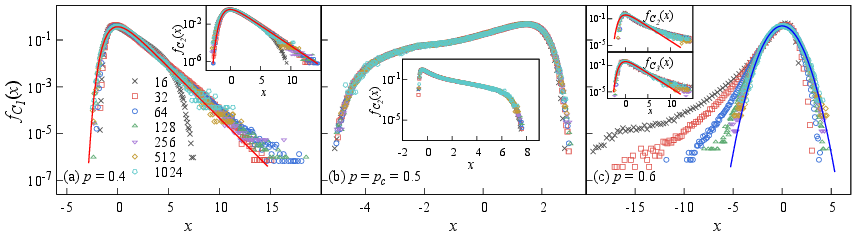}
\caption{  The probability distribution $f_{C_1}(x)$ of the largest-cluster size $\mathcal{C}_1$ for various side length $L$ and probability $p$. Here, the data are collected from BP on square lattice. In (a) and (b), $x$ is defined as Eq.~(\ref{eq03}), and $x =(\mathcal{C}_1 - C_1)/\sigma(\mathcal{C}_1)$ in (c). 
With this rescaling, if $\mathcal{C}_1$ follows a Gumbel/Gaussian distribution, $f_{C_1}(x)$ matches the standard Gumbel/Gaussian distribution, indicated by the red/blue line. For the same reason, 
the arguments $x$ for the size distributions of the second and third largest clusters, $f_{C_2}(x)$ and $f_{C_3}(x)$, in the insets 
of (a) and (c) are also defined as Eq.~(\ref{eq03}). (a) In the subcritical phase, both $f_{C_1}(x)$ and $f_{C_2}(x)$ (see inset) 
asymptotically approaches a Gumbel distribution. (b) At the percolation threshold $p_c$, $f_{C_1}(x)$ and $f_{C_2}(x)$ (see inset) are different, 
and clearly not a Gumbel distribution. (c) In the supercritical phase, $f_{C_1}(x)$ asymptotically approaches a Gaussian distribution. 
The inset shows that $f_{C_2}(x)$ and $f_{C_3}(x)$ asymptotically approach the Gumbel distribution.}  \label{fig2}
\end{figure*}

Extreme-value theory predicts that the maximum of a large number of independent and identically distributed random variables follow the Gumbel distribution. To test whether $f_{C_1}(x)$ conforms to the Gumbel distribution, we define $x$ as (see Appendix~A)
\begin{equation}
x \equiv \pi\frac{\mathcal{C}_1-C_1}{\sqrt{6} \sigma(\mathcal{C}_1)} +\gamma,
\label{eq03}
\end{equation} 
where $\mathcal{C}_1$ refers to the largest-cluster size in a single realization, $C_1\equiv\langle\mathcal{C}_1 \rangle$ is its mean, $\sigma(\mathcal{C}_1)$ is its standard deviation, and $\gamma \approx 0.5772$ is the Euler constant. With this rescaling, if $\mathcal{C}_1$ follows the Gumbel distribution, the probability distribution $f_{C_1}(x)$ should match the standard Gumbel distribution. To validate this, we simulate BP on square lattices with side length $L$ under the conventional ensemble and plot $f_{C_1}(x)$ for various $L$ and $p$ in Fig.~\ref{fig2}.

In the subcritical phase ($p\ll p_c$), Fig.~\ref{fig2}(a) illustrates that $f_{C_1}(x)$ asymptotically converges to the Gumbel distribution as $L$ increases. This observation indicates that $\mathcal{C}_1$ asymptotically follows extreme-value statistics in the subcritical phase. Such behavior is reasonable, as for $p \ll p_c$, the system only forms small clusters of similar sizes. The finite correlation length ensures that these clusters are effectively uncorrelated in sufficiently large systems and can be treated as independent random variables. Consequently, $f_{C_1}(x)$ adheres to the Gumbel distribution for large systems. A similar trend is observed for the second-largest cluster, as shown in the inset of Fig.~\ref{fig2}(a), which presents the probability distribution $f_{C_2}(x)$.

As $p$ approaches $p_c$, the correlation length diverges, introducing correlations between clusters. Consequently, the distribution $f_{C_1}(x)$ deviates significantly from the Gumbel distribution at $p=p_c$, as shown in Fig.~\ref{fig2}(b). Similarly, the size distribution $f_{C_2}(x)$ for the second-largest cluster also deviates from the Gumbel distribution (see inset). Additionally, the distributions $f_{C_1}(x)$ and $f_{C_2}(x)$ differ markedly in this regime.

In the supercritical phase ($p\gg p_c$), the largest cluster corresponds to the unique percolating cluster ($\sim V$), while the other clusters are significantly smaller ($\ll V$). Consequently, $\mathcal{C}_1$ is no long the maximum value selected from a set of random variables, leading to a clear deviation from the Gumbel distribution. Instead, as shown in Fig.~\ref{fig2}(c), $f_{C_1}(x)$ asymptotically converges to a Gaussian distribution, indicating that $\mathcal{C}_1$ obeys the central limit theorem for sufficiently large systems. Furthermore, the remaining clusters behave similarly to those in the subcritical phase. As a result, $f_{C_2}(x)$ and $f_{C_3}(x)$ again asymptotically follow the Gumbel distribution, as shown in the inset of Fig.~\ref{fig2}(c).

\begin{figure*}
\centering
    \includegraphics[width=\linewidth]{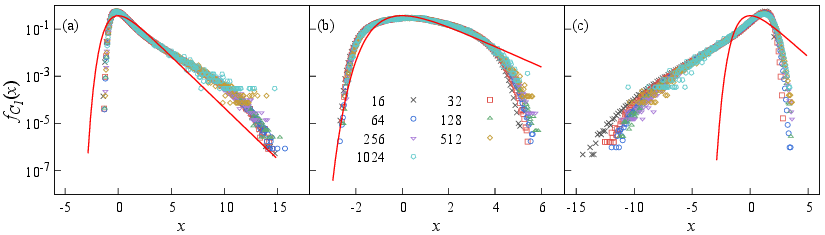}
    \caption{  The probability distribution $f_{C_1}(x)$ of the largest-cluster size $\mathcal{C}_1$ in the event-based ensemble for various side length $L$. Here, the data are collected from BP on square lattices, and $\mathcal{C}_1$ are rescaled as $x$ defined in Eq.~(\ref{eq03}). The sample positions are (a) $p = p_L - a L^{-1/\nu} $, (b) $p = p_L$, and (c) $p = p_L + a L^{-1/\nu}$, with $a = 1$ for simplicity. All these distributions are deviated from the standard Gumbel distribution, indicated by the red lines.}  \label{fig3}
\end{figure*}

The critical region can also be identified by a dynamic pseudocritical point $p_L$, i.e., in the event-based ensemble. In Fig.~\ref{fig3}(b), we present the probability density function $f_{C_1}(x)$ sampled at $p_L$. Notably, this distribution does not follow the Gumbel distribution, similar to the behavior observed at $p_c$. Moreover, the form of $f_{C_1}(x)$ differs between $p_L$ and $p_c$. In fact, within the critical finite-size scaling window $\mathcal{O}(L^{-1/\nu})$, the shape of $f_{C_1}(x)$ depends on the sample position $p$. In Figs.~\ref{fig3}(a) and (c), we illustrate $f_{C_1}(x)$ sampled at different positions in the critical finite-size scaling window, i.e., $p=p_L\pm aL^{-1/\nu}$ with $a=1$. The results show that all the distributions deviate from the Gumbel form and vary with the sample position $p$. The simulation results for some other models also suggest the deviation of $f_{C_1}(x)$ from the Gumbel distribution at pseudocritical points~\cite{feshanjerdi2024}.

Furthermore, the definition of $p_L$ is not unique. In addition to the maximum gap used in this paper, one could define $p_L$ based on the maximum point of the second moment of the cluster-size distribution, or the size of the second largest cluster. These pseudocritical points, defined in various ways, are all distributed around $p_c$ within the critical finite-size scaling window $\mathcal{O}(L^{-1/\nu})$. Consequently, as indicated by Fig.~\ref{fig3}, the distribution $f_{C_1}(x)$ at $p_L$ should vary with the definition of $p_L$. This lack of universality contrasts with the critical point $p_c$, where the distribution $f_{C_1}(x)$ is independent of lattice types and whether site or bond percolation is considered.

\begin{figure}
\centering
\includegraphics[width=\linewidth]{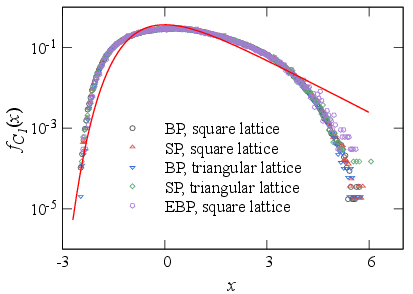}
\caption{ The probability distribution $f_{C_1}(x)$ in the event-based ensemble for 
different percolation models at pseudocritical point $p_L$. All cases have a side length of $L=1024$. 
The red line represents the standard Gumbel distribution.}  \label{fig4}
\end{figure}

However, for a given definition of $p_L$, as illustrated in Fig.~\ref{fig4}, the size distribution of the largest cluster $f_{C_1}(x)$ at $p_L$ for different 2D lattices -- such as BP an SP on square and triangular lattices, and EBP on square lattices -- collapses onto a single curve. This observation suggests a quasi-universality in the event-based ensemble for a specific definition of $p_L$.

\subsection{Dimensionless quantities}

At criticality, in addition to the universal distribution of observables, several dimensionless quantities take universal values across different lattices and for both site and bond percolation. For example, the wrapping probabilities at $p_c$ exhibit the universal relation $R_2 = R_0$ for planar lattices. Consequently, the critical polynomial has the universal value $P_B \equiv R_2 - R_0 = 0$ at $p_c$, with $R_2 = R_0 = 0.30952628$~\cite{ZIFF199917,PhysRevLett.85.4104} for square lattices under periodic boundary conditions. Moreover, for BP on square lattices with periodic boundary conditions, the Binder cumulants are $B_2=0.96017(1)$~\cite{hu2012percolation} and $B_4=0.87048(5)$~\cite{PhysRevB.71.144303, PhysRevE.71.016117}.  

\begin{table} 
\centering
    \tabcolsep=0.16 cm
    \caption{ Fitting results for the critical polynomial $P_B$ using the ansatz Eq.~(\ref{eq05}) with  $a_2=0$ and $y_2 = 0$. Data without 
    errors mean that they are fixed in the fit.}
    \begin{tabular}{clllll}
        \hline\hline
                       & $L_{\text{min}}$ & $O_{\infty}$   & $y_1$      & $a_{1}$        &  $\chi^{2}/DF$   \\
        \hline
        \multirow{6}{*}{\shortstack{BP on \\ square \\ lattice}} 
                                    &  16  &  -0.7395(1)   &  0.84(2)  &  -0.151(9)  &  3.98/6   \\
                                    &  32  &  -0.7393(2)   &  0.80(5)  &  -0.13(2)   &  3.55/5   \\
                                    &  64  &  -0.7390(4)   &  0.7(1)   &  -0.08(3)   &  2.47/4   \\
                                    &  16  &  -0.73927(7)  &  0.8      &  -0.136(1)  &  5.83/7   \\
                                    &  32  &  -0.73934(7)  &  0.8      &  -0.133(2)  &  3.55/6   \\
                                    &  64  &  -0.7394(1)   &  0.8      &  -0.130(5)  &  3.31/5   \\
        \hline
        \multirow{6}{*}{\shortstack{SP on \\ square \\ lattice}}   
                                    &  16  &  -0.73943(8)  &  1.41(1)  &  -1.65(7)   &  3.75/6   \\
                                    &  32  &  -0.7393(1)   &  1.35(4)  &  -1.3(2)    &  2.31/5   \\
                                    &  64  &  -0.73947(9)  &  1.53(9)  &  -3(1)      &  1.04/4   \\
                                    &  16  &  -0.73940(6)  &  1.4      &  -1.617(6)  &  3.91/7   \\
                                    &  32  &  -0.73944(6)  &  1.4      &  -1.60(2)   &  3.31/6   \\
                                    &  64  &  -0.73936(6)  &  1.4      &  -1.68(4)   &  1.59/5  \\
        \hline
        \multirow{4}{*}{\shortstack{BP on \\ triangular \\ lattice}}   
                                    & 32   &  -0.7527(2)   &  0.71(3)  &  -0.10(1)   &  1.02/5    \\
                                    & 64   &  -0.7526(3)   &  0.68(8)  &  -0.09(2)   &  0.95/4   \\
                                    & 32   &  -0.75263(6)  &  0.7      &  -0.100(1)  &  1.06/6   \\
                                    & 64   &  -0.75267(8)  &  0.7      &  -0.098(2)  &  0.97/5   \\
        \hline
        \multirow{4}{*}{\shortstack{SP on \\ triangular \\ lattice}}   
                                   &  32  &  -0.7526(2)   &  1.5(1) &  -1.8(6)     &  6.68/5   \\
                                   &  64  &  -0.7525(4)   &  1.4(4) &  -1(2)       &  6.61/4   \\
                                   &  32  &  -0.7526(1)   &  1.5    &  -1.80(4)    &  6.68/6   \\
                                   &  64  &  -0.7526(2)   &  1.5    &  -1.8(1)     &  6.67/5   \\
        \hline
        \multirow{4}{*}{\shortstack{EBP on \\ square \\ lattice}} 
                                    &  128 &  -0.7370(4)   &  1.13(4)  &  5(1)       &  0.41/3   \\
                                    &  256 &  -0.7371(7)   &  1.1(2)   &  5(4)       &  0.40/2   \\
                                    &  128 &  -0.7372(1)   &  1.1      &  4.64(3)    &  0.48/4   \\
                                    &  256 &  -0.7371(2)   &  1.1      &  4.5(1)     &  0.40/3   \\
        \hline\hline
    \end{tabular} \label{tab1}
\end{table}

\begin{figure}
\centering
    \includegraphics[width=\linewidth]{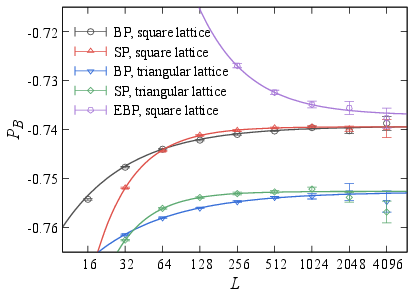}  
    \caption{Asymptotic behaviors of $P_B$ in the event-based ensemble for different  
    percolation models at $p_L$. The curves are obtained by fitting results in Table~\ref{tab1}. 
    Unlike the case at $p_c$, where $P_B=0$ universally holds, at $p_L$, the value of $P_B$
    is nonzero and no longer universal across different lattices.}  \label{fig5}
\end{figure}

As shown in Fig.~\ref{fig5}, at the pseudocritical point $p_L$, the critical polynomial $P_B$ also asymptotically approaches a constant for large systems. However, unlike at $p_c$, it does not exhibit a universal value across different lattices. To determine these values, we perform least-squares fits of the $P_B$ data using the finite-size scaling ansatz:  
\begin{equation}
    O(L)=O_{\infty}+a_{1}L^{-y_{1}}+a_{2}L^{-y_{2}},
    \label{eq05}
\end{equation}
where $O_{\infty}$ is the asymptotic value for infinite lattices, $y_1$ and $y_2$ are the correction exponents, and $a_1$ and $a_2$ are nonuniversal amplitudes. We set a lower cutoff of the side length $L_{\text{min}}$ on the data and monitor the residual $\chi^2$ as $L_{\text{min}}$ increased. When subsequent increases in $L_{\text{min}}$ no longer reduced $\chi^2$ by more than one unit per degree of freedom (DF), the fits were deemed stable. The fit results are summarized in Table~\ref{tab1}, showing that $P_B$ takes negative values because $p_L<p_c$ in 2D, leading to $R_0>R_2$.  

\begin{table} 
\centering
    \tabcolsep=0.16 cm
    \caption{ Fitting results for the wrapping probability $R_0$ 
    using the ansatz Eq.~(\ref{eq05}) with  $a_2=0$ and $y_2 = 0$. 
    Data without errors mean that they are fixed in the fit.}
        \begin{tabular}{clllll}
            \hline\hline
                                & $L_{\text{min}}$ & $O_{\infty}$   & $y_1$      & $a_{1}$        &  $\chi^{2}/DF$   \\
            \hline
            \multirow{6}{*}{\shortstack{BP on \\ square \\ lattice}}   
                                        &  16  &  0.7750(1)   &  0.82(3) &  0.099(8) &  5.35/6   \\
                                        &  32  &  0.7749(2)   &  0.81(7) &  0.09(2)  &  5.27/5   \\
                                        &  64  &  0.7746(4)   &  0.7(2)  &  0.05(3)  &  4.13/4   \\
                                        &  16  &  0.77487(6)  &  0.8     &  0.093(1) &  5.86/7   \\
                                        &  32  &  0.77490(7)  &  0.8     &  0.092(2) &  5.28/6   \\
                                        &  64  &  0.7749(1)   &  0.8     &  0.089(5) &  4.98/5   \\
            \hline
            \multirow{6}{*}{\shortstack{SP on \\ square \\ lattice}}  
                                        &  16  &  0.77492(6)  &  1.42(1) &  1.32(5)  &  2.85/6   \\
                                        &  32  &  0.77484(6)  &  1.36(3) &  1.1(1)   &  1.52/5   \\
                                        &  64  &  0.77492(7)  &  1.49(9) &  1.8(7)   &  0.94/4   \\
                                        &  16  &  0.77485(5)  &  1.4     &  1.233(5) &  4.29/7   \\
                                        &  32  &  0.77490(4)  &  1.4     &  1.21(1)  &  1.99/6   \\
                                        &  64  &  0.77486(4)  &  1.4     &  1.25(3)  &  1.19/5   \\
            \hline
            \multirow{5}{*}{\shortstack{BP on \\ triangular \\ lattice}}  
                                        &  32  &  0.7921(3)   &  0.69(6) &  0.07(1)  &  3.18/5   \\   
                                        &  64  &  0.7920(5)   &  0.6(2)  &  0.06(3)  &  3.07/4   \\
                                        &  32  &  0.79217(8)  &  0.7     &  0.070(1) &  3.20/6   \\
                                        &  64  &  0.7922(1)   &  0.7     &  0.070(3) &  3.20/5   \\
            \hline
            \multirow{5}{*}{\shortstack{SP on \\ triangular \\ lattice}}  
                                       &  32  &  0.7921(1)  &  1.45(9) &  1.1(3)    &  4.48/5   \\
                                       &  64  &  0.7921(2)  &  1.4(3)  &  1(1)    &  4.40/4   \\
                                       &  32  &  0.79222(8) &  1.5     &  1.33(3)   &  4.82/6   \\
                                       &  64  &  0.7922(1)  &  1.5     &  1.37(9)   &  4.61/5   \\
            \hline
            \multirow{4}{*}{\shortstack{EBP on \\ square \\ lattice}} 
                                        &  128 &  0.7733(3)   &  1.15(5) &  -3.9(9)  &  0.39/3   \\
                                        &  256 &  0.7733(6)   &  1.1(2)  &  -4(4)    &  0.39/2   \\
                                        &  128 &  0.7735(1)   &  1.1     &  -3.14(3) &  0.51/4   \\
                                        &  256 &  0.7734(2)   &  1.1     &  -3.0(1)  &  0.41/3   \\
            \hline\hline
        \end{tabular} \label{tab2}
\end{table}

From Table~\ref{tab1}, we conclude that $P_B = -0.7390(6)$ and $P_B = -0.7394(2)$ for BP and SP on square lattices, respectively. These two values are consistent within their error bars, indicating that $P_B$ is independent of bond or site percolation for a given lattice. The data obtained from the triangular lattice also support this conclusion; however, the value, $P_B = -0.7527(5)$ is different from that on square lattices. For EBP, it also exhibits a differentiated value, $P_B = -0.7371(7)$. The differences between these values are significantly larger than their error bars, confirming that $P_B$ at $p_L$ is not universal across different lattices. In summary, at the pseudocritical point $p_L$, the critical polynomial $P_B$ exhibits quasi-universal behavior: it depends on the lattice type but is independent of whether bond or site percolation is considered.

Moreover, by fitting the values of $R_0$ at $p_L$ to the finite-size scaling ansatz in Eq.~(\ref{eq05}), we also determine $R_0$ for various percolation systems, as listed in Table~\ref{tab2}. These results indicate that although the value of $R_0$ differs at $p_c$ and $p_L$, the universal behavior remains consistent, as $R_0$ does not depend on whether bond or site percolation is used, but does vary with the lattice type (boundary).  

\begin{table*} 
    \tabcolsep=0.6 cm
    \caption{Fitting results for the second order Binder cumulant $B_2$ using 
    the ansatz Eq.~(\ref{eq05}). Data without errors mean that they are fixed in the fit.}
        \begin{tabular}{clllllll}
            \hline\hline
                                        & $L_{\text{min}}$ & $O_{\infty}$   & $y_1 $      & $a_{1}$       & $y_2$     & $a_{2}$      & $ \chi^{2}/DF$   \\
            \hline
            \multirow{6}{*}{BP}         
                                        &  16   &  0.88617(4)  &  0.699(7)  &  0.061(1)   &  -        &  -         &  1.70/6   \\
                                        &  32   &  0.88624(7)  &  0.71(2)   &  0.064(4)   &  -        &  -         &  2.98/5   \\
                                        &  16   &  0.88619(2)  &  0.7       &  0.0612(2)  &  -        &   -        &  3.33/7   \\
                                        &  32   &  0.88617(2)  &  0.7       &  0.0614(3)  &  -        &  -         &  1.68/6   \\
                                        &  64   &  0.88617(3)  &  0.7       &  0.0615(7)  &  -        &  -         &  1.67/5   \\
                                        &  128  &  0.88617(5)  &  0.7       &  0.061(2)   &  -        &  -         &  1.67/4   \\
            \hline
            \multirow{3}{*}{SP}         
                                        &  16   &  0.8862(1) &  0.8(3) &  0.03(5) &  1.5(3) &  -0.6(1) &  0.98/4   \\
                                        &  32   &  0.88613(9) &  0.5(1) &  0.006(2) &  2     &  -1.6(1) &  0.54/4   \\
                                        &  32   &  0.88626(5) &  0.9(1) &  0.05(4) &  1.5  &  -0.6(1) &  0.96/4   \\
            \hline
            \multirow{4}{*}{EBP}        
                                        &  128  &  0.8855(3)   &  0.97(3)   &  -2.5(3)    &  -        &  -         &  1.75/3   \\
                                        &  256  &  0.8846(4)   &  1.13(9)   &  -5(2)      &  -        &  -         &  0.61/2   \\
                                        &  128  &  0.88518(9)  &  1         &  -2.78(1)   &  -        &  -         &  2.47/4   \\
                                        &  256  &  0.8854(1)   &  1         &  -2.85(4)   &  -        &  -         &  1.32/3   \\
            \hline\hline
        \end{tabular}  \label{tab3}
\end{table*}

In addition to the wrapping probabilities, we analyze the Binder cumulants at the pseudocritical point $p_L$ for several percolation models. The fitting results for $B_2$ and $B_4$ are presented in Table~\ref{tab3} and Table~\ref{tab4}, respectively. By comparing the fitting results for different $L_{\text{min}}$, we estimate the final values as $B_2=0.88617(5)$ and $B_4 = 0.5148(3)$ for BP on square lattices, $B_2 = 0.8860(2)$ and $B_4=0.51495(4)$ for SP on square lattices, and $B_2=0.8850(8)$ and $B_4=0.5150(8)$ for EBP. When compared to the values at $p_c$ ($B_2 = 0.96017(1)$ and $B_4=0.87048(5)$), the values of $B_2$ and $B_4$ at $p_L$ are significantly smaller. Nevertheless, within the error bars, these fit results further support the quasi-universality at $p_L$, consistent with the findings for wrapping probabilities.

\begin{table} 
\centering
    \tabcolsep=0.16 cm
    \caption{ Fitting results for the fourth order Binder 
    cumulant $B_4$ using the ansatz Eq.~(\ref{eq05}) with  $a_2=0$ and $y_2 = 0$. 
    Data without errors mean that they are fixed in the fit.}
        \begin{tabular}{clllll}
            \hline\hline
                                        & $L_{\text{min}} $& $O_{\infty}$   & $y_1$      & $a_{1}$        &$ \chi^{2}/DF$   \\
            \hline
            \multirow{3}{*}{BP}  
                                        &  64   &  0.5148(3)   &  0.6(2)   &  0.02(1)       &  5.64/4  \\
                                        &  128  &  0.5148(2)   &  0.6      &  0.019(2)       &  5.11/4  \\
                                        &  256  &  0.5149(2)   &  0.6      &  0.015(6)       &  4.18/3  \\
            \hline
            \multirow{3}{*}{SP}         
                                        &  32   &  0.51495(3)  &  2.5(1)   &  -19(9)         &  3.66/5  \\
                                        &  32   &  0.51496(2)  &  2.5      &  -16.7(2)       &  3.71/6 \\
                                        &  64   &  0.51496(3)  &  2.5      &  -17(2)         &  3.71/5  \\
            \hline
            \multirow{4}{*}{EBP} 
                                        &  128  &  0.5144(2)   &  1.12(2)  &  -4.3(4)        &  0.57/3  \\
                                        &  128  &  0.5155(2)   &  1.0      &  -2.50(3)       &  7.59/4  \\
                                        &  128  &  0.51459(6)  &  1.1      &  -3.87(1)       &  0.82/4  \\
                                        &  256  &  0.5145(1)   &  1.1      &  -3.83(6)       &  0.70/3  \\
            \hline\hline
        \end{tabular} \label{tab4}
\end{table}

\section{Discussion}
\label{sec-diss}

In summary, we investigate the universality of the distribution for the largest-cluster size and various dimensionless quantities in 2D percolation models within the event-based ensemble. Our numerical results reveal that the size distribution of the largest cluster asymptotically follows a Gumbel distribution in the subcritical phase and a Gaussian distribution in the supercritical phase. Within the critical finite-size scaling window, the distribution is neither Gumbel nor Gaussian, due to correlations induced by the diverging correlation length. Instead, the distribution varies with the sample position in the critical finite-size scaling window, highlighting the nonuniversality of the distribution of the largest-cluster size at pseudocritical points of different definitions. However, for a given pseudocritical point definition, this distribution remains universal across different lattices and bond or site percolation models, indicating a quasi-universality. For the wrapping probabilities and the critical polynomial, our numerical results also support a quasi-universality -- while their values depend on the lattice type, they are independent of whether bond or site percolation is considered. Similarly, the Binder cumulants also exhibit this quasi-universal behavior.  

Our findings suggest that the strict universality observed at the critical point is broken at the pseudocritical point, yet a quasi-universality persists. Recent studies on explosive percolation and rigidity percolation further indicate that quantities extracted at dynamic pseudocritical points tend to be less noisy and more reliable than those extracted at the critical point~\cite{PhysRevLett.130.147101,Li2024,lu2024}. These findings underscore subtle differences in critical behavior across various pseudocritical points, emphasizing that the choice of pseudocritical points may significantly influence the accuracy of scaling behavior analysis in certain models.  

\section*{Acknowledgment}
The authors acknowledge helpful discussions with Jingfang Fan. The research was supported by the National Natural Science Foundation of China under Grant No.~12275263, the Innovation Program for Quantum Science and Technology under Grant No.~2021ZD0301900, and Natural Science Foundation of Fujian province of China under Grant No.~2023J02032.

\appendix
\section*{Appendix A: Gumbel distribution}
\label{sec-app}

Let $F(x)$ denote the cumulative distribution function of the random variable $X$. For the Type I Gumbel distribution~\cite{beirlant2006statistics}, the cumulative distribution function is given by  
\begin{equation}
F(x;\mu,\beta) = e^{-e^{-\frac{x-\mu}{\beta}}},    \label{eq01}
\end{equation}  
where $\mu$ is the location parameter and $\beta$ is the scale parameter. The corresponding probability density function, $f(x)$, is 
\begin{equation}
f(x;\mu,\beta) = F^\prime(x;\mu,\beta) = \frac{1}{\beta} e^{-\frac{x-\mu}{\beta} -e^{-\frac{x-\mu}{\beta}}}.    \label{eq02}
\end{equation}  
The standard Gumbel distribution is a special case where $\mu=0$ and $\beta = 1$, yielding  
\begin{equation}
f(x) = e^{-x - e^{-x}}.
\end{equation}  

From Eq.~(\ref{eq02}), the mean and standard deviation of a Gumbel distribution are given by  
\begin{align}
\langle X \rangle &= \mu + \gamma \beta, \\
\sigma(X) &= \frac{\pi\beta}{\sqrt{6}},
\end{align}  
where $\gamma \approx 0.5772$ is the Euler constant. This implies that the standard Gumbel distribution has a mean of $\gamma$ and a standard deviation of $\pi/\sqrt{6}$.

For any random variable $X$ following a Gumbel distribution Eq.~(\ref{eq02}), it is easy to verify that the rescaled random variable  
\begin{equation}
X' \equiv \frac{X - \langle X \rangle}{\sigma(X)} \frac{\pi}{\sqrt{6}} + \gamma
\end{equation}  
should conform to the standard Gumbel distribution.

\bibliography{ref.bib}

\begin{thebibliography}{44}%
\makeatletter
\providecommand \@ifxundefined [1]{%
 \@ifx{#1\undefined}
}%
\providecommand \@ifnum [1]{%
 \ifnum #1\expandafter \@firstoftwo
 \else \expandafter \@secondoftwo
 \fi
}%
\providecommand \@ifx [1]{%
 \ifx #1\expandafter \@firstoftwo
 \else \expandafter \@secondoftwo
 \fi
}%
\providecommand \natexlab [1]{#1}%
\providecommand \enquote  [1]{``#1''}%
\providecommand \bibnamefont  [1]{#1}%
\providecommand \bibfnamefont [1]{#1}%
\providecommand \citenamefont [1]{#1}%
\providecommand \href@noop [0]{\@secondoftwo}%
\providecommand \href [0]{\begingroup \@sanitize@url \@href}%
\providecommand \@href[1]{\@@startlink{#1}\@@href}%
\providecommand \@@href[1]{\endgroup#1\@@endlink}%
\providecommand \@sanitize@url [0]{\catcode `\\12\catcode `\$12\catcode `\&12\catcode `\#12\catcode `\^12\catcode `\_12\catcode `\%12\relax}%
\providecommand \@@startlink[1]{}%
\providecommand \@@endlink[0]{}%
\providecommand \url  [0]{\begingroup\@sanitize@url \@url }%
\providecommand \@url [1]{\endgroup\@href {#1}{\urlprefix }}%
\providecommand \urlprefix  [0]{URL }%
\providecommand \Eprint [0]{\href }%
\providecommand \doibase [0]{https://doi.org/}%
\providecommand \selectlanguage [0]{\@gobble}%
\providecommand \bibinfo  [0]{\@secondoftwo}%
\providecommand \bibfield  [0]{\@secondoftwo}%
\providecommand \translation [1]{[#1]}%
\providecommand \BibitemOpen [0]{}%
\providecommand \bibitemStop [0]{}%
\providecommand \bibitemNoStop [0]{.\EOS\space}%
\providecommand \EOS [0]{\spacefactor3000\relax}%
\providecommand \BibitemShut  [1]{\csname bibitem#1\endcsname}%
\let\auto@bib@innerbib\@empty
\bibitem [{\citenamefont {Stauffer}\ and\ \citenamefont {Aharony}(1991)}]{stauffer1991}%
  \BibitemOpen
  \bibfield  {author} {\bibinfo {author} {\bibfnamefont {D.}~\bibnamefont {Stauffer}}\ and\ \bibinfo {author} {\bibfnamefont {A.}~\bibnamefont {Aharony}},\ }\href@noop {} {\emph {\bibinfo {title} {Introduction to percolation theory}}}\ (\bibinfo  {publisher} {Taylor \& Francis},\ \bibinfo {year} {1991})\BibitemShut {NoStop}%
\bibitem [{\citenamefont {Kim}\ \emph {et~al.}(2016)\citenamefont {Kim}, \citenamefont {Choi}, \citenamefont {Oh}, \citenamefont {Byun}, \citenamefont {Kim}, \citenamefont {Lee}, \citenamefont {Lee},\ and\ \citenamefont {Hong}}]{kim2016revisit}%
  \BibitemOpen
  \bibfield  {author} {\bibinfo {author} {\bibfnamefont {S.}~\bibnamefont {Kim}}, \bibinfo {author} {\bibfnamefont {S.}~\bibnamefont {Choi}}, \bibinfo {author} {\bibfnamefont {E.}~\bibnamefont {Oh}}, \bibinfo {author} {\bibfnamefont {J.}~\bibnamefont {Byun}}, \bibinfo {author} {\bibfnamefont {H.}~\bibnamefont {Kim}}, \bibinfo {author} {\bibfnamefont {B.}~\bibnamefont {Lee}}, \bibinfo {author} {\bibfnamefont {S.}~\bibnamefont {Lee}},\ and\ \bibinfo {author} {\bibfnamefont {Y.}~\bibnamefont {Hong}},\ }\bibfield  {title} {\bibinfo {title} {Revisit to three-dimensional percolation theory: Accurate analysis for highly stretchable conductive composite materials},\ }\href {https://doi.org/10.1038/srep34632} {\bibfield  {journal} {\bibinfo  {journal} {Sci. Rep.}\ }\textbf {\bibinfo {volume} {6}},\ \bibinfo {pages} {34632} (\bibinfo {year} {2016})}\BibitemShut {NoStop}%
\bibitem [{\citenamefont {Li}\ \emph {et~al.}(2021)\citenamefont {Li}, \citenamefont {Liu}, \citenamefont {Lü}, \citenamefont {Hu}, \citenamefont {Xu},\ and\ \citenamefont {Zhang}}]{LI20211}%
  \BibitemOpen
  \bibfield  {author} {\bibinfo {author} {\bibfnamefont {M.}~\bibnamefont {Li}}, \bibinfo {author} {\bibfnamefont {R.-R.}\ \bibnamefont {Liu}}, \bibinfo {author} {\bibfnamefont {L.}~\bibnamefont {Lü}}, \bibinfo {author} {\bibfnamefont {M.-B.}\ \bibnamefont {Hu}}, \bibinfo {author} {\bibfnamefont {S.}~\bibnamefont {Xu}},\ and\ \bibinfo {author} {\bibfnamefont {Y.-C.}\ \bibnamefont {Zhang}},\ }\bibfield  {title} {\bibinfo {title} {Percolation on complex networks: Theory and application},\ }\href {https://doi.org/https://doi.org/10.1016/j.physrep.2020.12.003} {\bibfield  {journal} {\bibinfo  {journal} {Phys. Rep.}\ }\textbf {\bibinfo {volume} {907}},\ \bibinfo {pages} {1} (\bibinfo {year} {2021})}\BibitemShut {NoStop}%
\bibitem [{\citenamefont {Ziff}(2021)}]{ziff2021percolation}%
  \BibitemOpen
  \bibfield  {author} {\bibinfo {author} {\bibfnamefont {R.~M.}\ \bibnamefont {Ziff}},\ }\bibfield  {title} {\bibinfo {title} {Percolation and the pandemic},\ }\href {https://doi.org/10.1016/j.physa.2020.125723} {\bibfield  {journal} {\bibinfo  {journal} {Physica A}\ }\textbf {\bibinfo {volume} {568}},\ \bibinfo {pages} {125723} (\bibinfo {year} {2021})}\BibitemShut {NoStop}%
\bibitem [{\citenamefont {Sykes}\ and\ \citenamefont {Essam}(1964)}]{10.1063/1.1704215}%
  \BibitemOpen
  \bibfield  {author} {\bibinfo {author} {\bibfnamefont {M.~F.}\ \bibnamefont {Sykes}}\ and\ \bibinfo {author} {\bibfnamefont {J.~W.}\ \bibnamefont {Essam}},\ }\bibfield  {title} {\bibinfo {title} {Exact critical percolation probabilities for site and bond problems in two dimensions},\ }\href {https://doi.org/10.1063/1.1704215} {\bibfield  {journal} {\bibinfo  {journal} {J. Math. Phys.}\ }\textbf {\bibinfo {volume} {5}},\ \bibinfo {pages} {1117} (\bibinfo {year} {1964})}\BibitemShut {NoStop}%
\bibitem [{\citenamefont {Lieb}(1967)}]{PhysRevLett.18.692}%
  \BibitemOpen
  \bibfield  {author} {\bibinfo {author} {\bibfnamefont {E.~H.}\ \bibnamefont {Lieb}},\ }\bibfield  {title} {\bibinfo {title} {Exact solution of the problem of the entropy of two-dimensional ice},\ }\href {https://doi.org/10.1103/PhysRevLett.18.692} {\bibfield  {journal} {\bibinfo  {journal} {Phys. Rev. Lett.}\ }\textbf {\bibinfo {volume} {18}},\ \bibinfo {pages} {692} (\bibinfo {year} {1967})}\BibitemShut {NoStop}%
\bibitem [{\citenamefont {Baxter}(1972)}]{BAXTER1972193}%
  \BibitemOpen
  \bibfield  {author} {\bibinfo {author} {\bibfnamefont {R.~J.}\ \bibnamefont {Baxter}},\ }\bibfield  {title} {\bibinfo {title} {Partition function of the eight-vertex lattice model},\ }\href {https://doi.org/10.1016/0003-4916(72)90335-1} {\bibfield  {journal} {\bibinfo  {journal} {Ann. Phys.}\ }\textbf {\bibinfo {volume} {70}},\ \bibinfo {pages} {193} (\bibinfo {year} {1972})}\BibitemShut {NoStop}%
\bibitem [{\citenamefont {Belavin}\ \emph {et~al.}(1984)\citenamefont {Belavin}, \citenamefont {Polyakov},\ and\ \citenamefont {Zamolodchikov}}]{BELAVIN1984333}%
  \BibitemOpen
  \bibfield  {author} {\bibinfo {author} {\bibfnamefont {A.}~\bibnamefont {Belavin}}, \bibinfo {author} {\bibfnamefont {A.}~\bibnamefont {Polyakov}},\ and\ \bibinfo {author} {\bibfnamefont {A.}~\bibnamefont {Zamolodchikov}},\ }\bibfield  {title} {\bibinfo {title} {Infinite conformal symmetry in two-dimensional quantum field theory},\ }\href {https://doi.org/10.1016/0550-3213(84)90052-X} {\bibfield  {journal} {\bibinfo  {journal} {Nucl. Phys. B}\ }\textbf {\bibinfo {volume} {241}},\ \bibinfo {pages} {333} (\bibinfo {year} {1984})}\BibitemShut {NoStop}%
\bibitem [{\citenamefont {Friedan}\ \emph {et~al.}(1984)\citenamefont {Friedan}, \citenamefont {Qiu},\ and\ \citenamefont {Shenker}}]{PhysRevLett.52.1575}%
  \BibitemOpen
  \bibfield  {author} {\bibinfo {author} {\bibfnamefont {D.}~\bibnamefont {Friedan}}, \bibinfo {author} {\bibfnamefont {Z.}~\bibnamefont {Qiu}},\ and\ \bibinfo {author} {\bibfnamefont {S.}~\bibnamefont {Shenker}},\ }\bibfield  {title} {\bibinfo {title} {Conformal invariance, unitarity, and critical exponents in two dimensions},\ }\href {https://doi.org/10.1103/PhysRevLett.52.1575} {\bibfield  {journal} {\bibinfo  {journal} {Phys. Rev. Lett.}\ }\textbf {\bibinfo {volume} {52}},\ \bibinfo {pages} {1575} (\bibinfo {year} {1984})}\BibitemShut {NoStop}%
\bibitem [{\citenamefont {Nienhuis}(1984)}]{nienhuis1984critical}%
  \BibitemOpen
  \bibfield  {author} {\bibinfo {author} {\bibfnamefont {B.}~\bibnamefont {Nienhuis}},\ }\bibfield  {title} {\bibinfo {title} {Critical behavior of two-dimensional spin models and charge asymmetry in the coulomb gas},\ }\href {https://doi.org/10.1007/BF01009437} {\bibfield  {journal} {\bibinfo  {journal} {J. Stat. Phys.}\ }\textbf {\bibinfo {volume} {34}},\ \bibinfo {pages} {731} (\bibinfo {year} {1984})}\BibitemShut {NoStop}%
\bibitem [{\citenamefont {Cardy}(1986)}]{CARDY1986219}%
  \BibitemOpen
  \bibfield  {author} {\bibinfo {author} {\bibfnamefont {J.~L.}\ \bibnamefont {Cardy}},\ }\bibfield  {title} {\bibinfo {title} {Conformal invariance and critical behavior},\ }\href {https://doi.org/https://doi.org/10.1016/0378-4371(86)90225-6} {\bibfield  {journal} {\bibinfo  {journal} {Physica A}\ }\textbf {\bibinfo {volume} {140}},\ \bibinfo {pages} {219} (\bibinfo {year} {1986})}\BibitemShut {NoStop}%
\bibitem [{\citenamefont {Cardy}(2005)}]{CARDY200581}%
  \BibitemOpen
  \bibfield  {author} {\bibinfo {author} {\bibfnamefont {J.}~\bibnamefont {Cardy}},\ }\bibfield  {title} {\bibinfo {title} {Sle for theoretical physicists},\ }\href {https://doi.org/10.1016/j.aop.2005.04.001} {\bibfield  {journal} {\bibinfo  {journal} {Ann. Phys.}\ }\textbf {\bibinfo {volume} {318}},\ \bibinfo {pages} {81} (\bibinfo {year} {2005})},\ \bibinfo {note} {special Issue}\BibitemShut {NoStop}%
\bibitem [{\citenamefont {Smirnov}\ and\ \citenamefont {Werner}(2001)}]{smirnov2001critical}%
  \BibitemOpen
  \bibfield  {author} {\bibinfo {author} {\bibfnamefont {S.}~\bibnamefont {Smirnov}}\ and\ \bibinfo {author} {\bibfnamefont {W.}~\bibnamefont {Werner}},\ }\bibfield  {title} {\bibinfo {title} {Critical exponents for two-dimensional percolation},\ }\href {https://doi.org/10.4310/mrl.2001.v8.n4.a1} {\bibfield  {journal} {\bibinfo  {journal} {Math. Res. Lett.}\ }\textbf {\bibinfo {volume} {8}},\ \bibinfo {pages} {729} (\bibinfo {year} {2001})}\BibitemShut {NoStop}%
\bibitem [{\citenamefont {Grassberger}\ \emph {et~al.}(2011)\citenamefont {Grassberger}, \citenamefont {Christensen}, \citenamefont {Bizhani}, \citenamefont {Son},\ and\ \citenamefont {Paczuski}}]{Grassberger2011}%
  \BibitemOpen
  \bibfield  {author} {\bibinfo {author} {\bibfnamefont {P.}~\bibnamefont {Grassberger}}, \bibinfo {author} {\bibfnamefont {C.}~\bibnamefont {Christensen}}, \bibinfo {author} {\bibfnamefont {G.}~\bibnamefont {Bizhani}}, \bibinfo {author} {\bibfnamefont {S.-W.}\ \bibnamefont {Son}},\ and\ \bibinfo {author} {\bibfnamefont {M.}~\bibnamefont {Paczuski}},\ }\bibfield  {title} {\bibinfo {title} {Explosive percolation is continuous, but with unusual finite size behavior},\ }\href {https://doi.org/10.1103/physrevlett.106.225701} {\bibfield  {journal} {\bibinfo  {journal} {Phys. Rev. Lett.}\ }\textbf {\bibinfo {volume} {106}},\ \bibinfo {pages} {225701} (\bibinfo {year} {2011})}\BibitemShut {NoStop}%
\bibitem [{\citenamefont {D’Souza}\ and\ \citenamefont {Nagler}(2015)}]{Souza2015}%
  \BibitemOpen
  \bibfield  {author} {\bibinfo {author} {\bibfnamefont {R.~M.}\ \bibnamefont {D’Souza}}\ and\ \bibinfo {author} {\bibfnamefont {J.}~\bibnamefont {Nagler}},\ }\bibfield  {title} {\bibinfo {title} {Anomalous critical and supercritical phenomena in explosive percolation},\ }\href {https://doi.org/10.1038/nphys3378} {\bibfield  {journal} {\bibinfo  {journal} {Nat. Phys.}\ }\textbf {\bibinfo {volume} {11}},\ \bibinfo {pages} {531} (\bibinfo {year} {2015})}\BibitemShut {NoStop}%
\bibitem [{\citenamefont {Achlioptas}\ \emph {et~al.}(2009)\citenamefont {Achlioptas}, \citenamefont {D’Souza},\ and\ \citenamefont {Spencer}}]{Achlioptas2009}%
  \BibitemOpen
  \bibfield  {author} {\bibinfo {author} {\bibfnamefont {D.}~\bibnamefont {Achlioptas}}, \bibinfo {author} {\bibfnamefont {R.~M.}\ \bibnamefont {D’Souza}},\ and\ \bibinfo {author} {\bibfnamefont {J.}~\bibnamefont {Spencer}},\ }\bibfield  {title} {\bibinfo {title} {Explosive percolation in random networks},\ }\href {https://doi.org/10.1126/science.1167782} {\bibfield  {journal} {\bibinfo  {journal} {Science}\ }\textbf {\bibinfo {volume} {323}},\ \bibinfo {pages} {1453} (\bibinfo {year} {2009})}\BibitemShut {NoStop}%
\bibitem [{\citenamefont {Friedman}\ and\ \citenamefont {Landsberg}(2009)}]{Friedman2009}%
  \BibitemOpen
  \bibfield  {author} {\bibinfo {author} {\bibfnamefont {E.~J.}\ \bibnamefont {Friedman}}\ and\ \bibinfo {author} {\bibfnamefont {A.~S.}\ \bibnamefont {Landsberg}},\ }\bibfield  {title} {\bibinfo {title} {Construction and analysis of random networks with explosive percolation},\ }\href {https://doi.org/10.1103/physrevlett.103.255701} {\bibfield  {journal} {\bibinfo  {journal} {Phys. Rev. Lett.}\ }\textbf {\bibinfo {volume} {103}},\ \bibinfo {pages} {255701} (\bibinfo {year} {2009})}\BibitemShut {NoStop}%
\bibitem [{\citenamefont {Ziff}(2009)}]{Ziff2009}%
  \BibitemOpen
  \bibfield  {author} {\bibinfo {author} {\bibfnamefont {R.~M.}\ \bibnamefont {Ziff}},\ }\bibfield  {title} {\bibinfo {title} {Explosive growth in biased dynamic percolation on two-dimensional regular lattice networks},\ }\href {https://doi.org/10.1103/physrevlett.103.045701} {\bibfield  {journal} {\bibinfo  {journal} {Phys. Rev. Lett.}\ }\textbf {\bibinfo {volume} {103}},\ \bibinfo {pages} {045701} (\bibinfo {year} {2009})}\BibitemShut {NoStop}%
\bibitem [{\citenamefont {D’Souza}\ and\ \citenamefont {Mitzenmacher}(2010)}]{Souza2010}%
  \BibitemOpen
  \bibfield  {author} {\bibinfo {author} {\bibfnamefont {R.~M.}\ \bibnamefont {D’Souza}}\ and\ \bibinfo {author} {\bibfnamefont {M.}~\bibnamefont {Mitzenmacher}},\ }\bibfield  {title} {\bibinfo {title} {Local cluster aggregation models of explosive percolation},\ }\href {https://doi.org/10.1103/physrevlett.104.195702} {\bibfield  {journal} {\bibinfo  {journal} {Phys. Rev. Lett.}\ }\textbf {\bibinfo {volume} {104}},\ \bibinfo {pages} {195702} (\bibinfo {year} {2010})}\BibitemShut {NoStop}%
\bibitem [{\citenamefont {Li}\ \emph {et~al.}(2023)\citenamefont {Li}, \citenamefont {Wang},\ and\ \citenamefont {Deng}}]{PhysRevLett.130.147101}%
  \BibitemOpen
  \bibfield  {author} {\bibinfo {author} {\bibfnamefont {M.}~\bibnamefont {Li}}, \bibinfo {author} {\bibfnamefont {J.}~\bibnamefont {Wang}},\ and\ \bibinfo {author} {\bibfnamefont {Y.}~\bibnamefont {Deng}},\ }\bibfield  {title} {\bibinfo {title} {Explosive percolation obeys standard finite-size scaling in an event-based ensemble},\ }\href {https://doi.org/10.1103/PhysRevLett.130.147101} {\bibfield  {journal} {\bibinfo  {journal} {Phys. Rev. Lett.}\ }\textbf {\bibinfo {volume} {130}},\ \bibinfo {pages} {147101} (\bibinfo {year} {2023})}\BibitemShut {NoStop}%
\bibitem [{\citenamefont {Li}\ \emph {et~al.}(2024{\natexlab{a}})\citenamefont {Li}, \citenamefont {Wang},\ and\ \citenamefont {Deng}}]{Li2024}%
  \BibitemOpen
  \bibfield  {author} {\bibinfo {author} {\bibfnamefont {M.}~\bibnamefont {Li}}, \bibinfo {author} {\bibfnamefont {J.}~\bibnamefont {Wang}},\ and\ \bibinfo {author} {\bibfnamefont {Y.}~\bibnamefont {Deng}},\ }\bibfield  {title} {\bibinfo {title} {Explosive percolation in finite dimensions},\ }\href {https://doi.org/10.1103/physrevresearch.6.033319} {\bibfield  {journal} {\bibinfo  {journal} {Phys. Rev. Research}\ }\textbf {\bibinfo {volume} {6}},\ \bibinfo {pages} {033319} (\bibinfo {year} {2024}{\natexlab{a}})}\BibitemShut {NoStop}%
\bibitem [{\citenamefont {Lu}\ \emph {et~al.}(2024{\natexlab{a}})\citenamefont {Lu}, \citenamefont {Fang}, \citenamefont {Zhou},\ and\ \citenamefont {Deng}}]{PhysRevE.110.044140}%
  \BibitemOpen
  \bibfield  {author} {\bibinfo {author} {\bibfnamefont {M.}~\bibnamefont {Lu}}, \bibinfo {author} {\bibfnamefont {S.}~\bibnamefont {Fang}}, \bibinfo {author} {\bibfnamefont {Z.}~\bibnamefont {Zhou}},\ and\ \bibinfo {author} {\bibfnamefont {Y.}~\bibnamefont {Deng}},\ }\bibfield  {title} {\bibinfo {title} {Interplay of the complete-graph and gaussian fixed-point asymptotics in finite-size scaling of percolation above the upper critical dimension},\ }\href {https://doi.org/10.1103/PhysRevE.110.044140} {\bibfield  {journal} {\bibinfo  {journal} {Phys. Rev. E}\ }\textbf {\bibinfo {volume} {110}},\ \bibinfo {pages} {044140} (\bibinfo {year} {2024}{\natexlab{a}})}\BibitemShut {NoStop}%
\bibitem [{\citenamefont {Kenna}\ and\ \citenamefont {Berche}(2017)}]{Kenna2017}%
  \BibitemOpen
  \bibfield  {author} {\bibinfo {author} {\bibfnamefont {R.}~\bibnamefont {Kenna}}\ and\ \bibinfo {author} {\bibfnamefont {B.}~\bibnamefont {Berche}},\ }\bibfield  {title} {\bibinfo {title} {Universal finite-size scaling for percolation theory in high dimensions},\ }\href {https://doi.org/10.1088/1751-8121/aa6bd5} {\bibfield  {journal} {\bibinfo  {journal} {J. Phys. A: Math. Theor.}\ }\textbf {\bibinfo {volume} {50}},\ \bibinfo {pages} {235001} (\bibinfo {year} {2017})}\BibitemShut {NoStop}%
\bibitem [{\citenamefont {Borgs}\ \emph {et~al.}(2005{\natexlab{a}})\citenamefont {Borgs}, \citenamefont {Chayes}, \citenamefont {van~der Hofstad}, \citenamefont {Slade},\ and\ \citenamefont {Spencer}}]{rsa.20051}%
  \BibitemOpen
  \bibfield  {author} {\bibinfo {author} {\bibfnamefont {C.}~\bibnamefont {Borgs}}, \bibinfo {author} {\bibfnamefont {J.~T.}\ \bibnamefont {Chayes}}, \bibinfo {author} {\bibfnamefont {R.}~\bibnamefont {van~der Hofstad}}, \bibinfo {author} {\bibfnamefont {G.}~\bibnamefont {Slade}},\ and\ \bibinfo {author} {\bibfnamefont {J.}~\bibnamefont {Spencer}},\ }\bibfield  {title} {\bibinfo {title} {Random subgraphs of finite graphs: I. the scaling window under the triangle condition},\ }\href {https://doi.org/10.1002/rsa.20051} {\bibfield  {journal} {\bibinfo  {journal} {Random Struct. Algorithms}\ }\textbf {\bibinfo {volume} {27}},\ \bibinfo {pages} {137} (\bibinfo {year} {2005}{\natexlab{a}})}\BibitemShut {NoStop}%
\bibitem [{\citenamefont {Borgs}\ \emph {et~al.}(2005{\natexlab{b}})\citenamefont {Borgs}, \citenamefont {Chayes}, \citenamefont {Slade}, \citenamefont {Spencer},\ and\ \citenamefont {van~der Hofstad}}]{ChristianRandom}%
  \BibitemOpen
  \bibfield  {author} {\bibinfo {author} {\bibfnamefont {C.}~\bibnamefont {Borgs}}, \bibinfo {author} {\bibfnamefont {J.~T.}\ \bibnamefont {Chayes}}, \bibinfo {author} {\bibfnamefont {G.}~\bibnamefont {Slade}}, \bibinfo {author} {\bibfnamefont {J.}~\bibnamefont {Spencer}},\ and\ \bibinfo {author} {\bibfnamefont {R.}~\bibnamefont {van~der Hofstad}},\ }\bibfield  {title} {\bibinfo {title} {Random subgraphs of finite graphs. ii. the lace expansion and the triangle condition},\ }\href {https://doi.org/10.1214/009117905000000260} {\bibfield  {journal} {\bibinfo  {journal} {Ann. Probab.}\ }\textbf {\bibinfo {volume} {33}},\ \bibinfo {pages} {1886 } (\bibinfo {year} {2005}{\natexlab{b}})}\BibitemShut {NoStop}%
\bibitem [{\citenamefont {Heydenreich}\ and\ \citenamefont {Van Der~Hofstad}(2007)}]{heydenreich2007random}%
  \BibitemOpen
  \bibfield  {author} {\bibinfo {author} {\bibfnamefont {M.}~\bibnamefont {Heydenreich}}\ and\ \bibinfo {author} {\bibfnamefont {R.}~\bibnamefont {Van Der~Hofstad}},\ }\bibfield  {title} {\bibinfo {title} {Random graph asymptotics on high-dimensional tori},\ }\href {https://doi.org/10.1007/s00220-006-0152-8} {\bibfield  {journal} {\bibinfo  {journal} {Commun. Math. Phys.}\ }\textbf {\bibinfo {volume} {270}},\ \bibinfo {pages} {335} (\bibinfo {year} {2007})}\BibitemShut {NoStop}%
\bibitem [{\citenamefont {Aizenman}(1997)}]{AIZENMAN1997551}%
  \BibitemOpen
  \bibfield  {author} {\bibinfo {author} {\bibfnamefont {M.}~\bibnamefont {Aizenman}},\ }\bibfield  {title} {\bibinfo {title} {On the number of incipient spanning clusters},\ }\href {https://doi.org/https://doi.org/10.1016/S0550-3213(96)00626-8} {\bibfield  {journal} {\bibinfo  {journal} {Nucl. Phys. B}\ }\textbf {\bibinfo {volume} {485}},\ \bibinfo {pages} {551} (\bibinfo {year} {1997})}\BibitemShut {NoStop}%
\bibitem [{\citenamefont {Li}\ \emph {et~al.}(2024{\natexlab{b}})\citenamefont {Li}, \citenamefont {Fang}, \citenamefont {Fan},\ and\ \citenamefont {Deng}}]{crossoverfinitesizescalingtheory}%
  \BibitemOpen
  \bibfield  {author} {\bibinfo {author} {\bibfnamefont {M.}~\bibnamefont {Li}}, \bibinfo {author} {\bibfnamefont {S.}~\bibnamefont {Fang}}, \bibinfo {author} {\bibfnamefont {J.}~\bibnamefont {Fan}},\ and\ \bibinfo {author} {\bibfnamefont {Y.}~\bibnamefont {Deng}},\ }\bibfield  {title} {\bibinfo {title} {Crossover finite-size scaling theory and its applications in percolation},\ }\href {https://arxiv.org/abs/2412.06228} {\bibfield  {journal} {\bibinfo  {journal} {arXiv preprint arXiv:2412.06228}\ } (\bibinfo {year} {2024}{\natexlab{b}})}\BibitemShut {NoStop}%
\bibitem [{\citenamefont {Fan}\ \emph {et~al.}(2020)\citenamefont {Fan}, \citenamefont {Meng}, \citenamefont {Liu}, \citenamefont {Saberi}, \citenamefont {Kurths},\ and\ \citenamefont {Nagler}}]{fan2020universal}%
  \BibitemOpen
  \bibfield  {author} {\bibinfo {author} {\bibfnamefont {J.}~\bibnamefont {Fan}}, \bibinfo {author} {\bibfnamefont {J.}~\bibnamefont {Meng}}, \bibinfo {author} {\bibfnamefont {Y.}~\bibnamefont {Liu}}, \bibinfo {author} {\bibfnamefont {A.~A.}\ \bibnamefont {Saberi}}, \bibinfo {author} {\bibfnamefont {J.}~\bibnamefont {Kurths}},\ and\ \bibinfo {author} {\bibfnamefont {J.}~\bibnamefont {Nagler}},\ }\bibfield  {title} {\bibinfo {title} {Universal gap scaling in percolation},\ }\href {https://doi.org/10.1038/s41567-019-0783-2} {\bibfield  {journal} {\bibinfo  {journal} {Nat. Phys.}\ }\textbf {\bibinfo {volume} {16}},\ \bibinfo {pages} {455} (\bibinfo {year} {2020})}\BibitemShut {NoStop}%
\bibitem [{\citenamefont {Langlands}\ \emph {et~al.}(1992)\citenamefont {Langlands}, \citenamefont {Pichet}, \citenamefont {Pouliot},\ and\ \citenamefont {Saint-Aubin}}]{langlands1992}%
  \BibitemOpen
  \bibfield  {author} {\bibinfo {author} {\bibfnamefont {R.~P.}\ \bibnamefont {Langlands}}, \bibinfo {author} {\bibfnamefont {C.}~\bibnamefont {Pichet}}, \bibinfo {author} {\bibfnamefont {P.}~\bibnamefont {Pouliot}},\ and\ \bibinfo {author} {\bibfnamefont {Y.}~\bibnamefont {Saint-Aubin}},\ }\bibfield  {title} {\bibinfo {title} {On the universality of crossing probabilities in two-dimensional percolation},\ }\href {https://doi.org/10.1007/BF01049720} {\bibfield  {journal} {\bibinfo  {journal} {J. Stat. Phys.}\ }\textbf {\bibinfo {volume} {67}},\ \bibinfo {pages} {553} (\bibinfo {year} {1992})}\BibitemShut {NoStop}%
\bibitem [{\citenamefont {Pinson}(1994)}]{pinson1994critical}%
  \BibitemOpen
  \bibfield  {author} {\bibinfo {author} {\bibfnamefont {H.~T.}\ \bibnamefont {Pinson}},\ }\bibfield  {title} {\bibinfo {title} {Critical percolation on the torus},\ }\href {https://doi.org/10.1007/BF02186762} {\bibfield  {journal} {\bibinfo  {journal} {J. Stat. Phys.}\ }\textbf {\bibinfo {volume} {75}},\ \bibinfo {pages} {1167} (\bibinfo {year} {1994})}\BibitemShut {NoStop}%
\bibitem [{\citenamefont {Ziff}\ \emph {et~al.}(1999)\citenamefont {Ziff}, \citenamefont {Lorenz},\ and\ \citenamefont {Kleban}}]{ZIFF199917}%
  \BibitemOpen
  \bibfield  {author} {\bibinfo {author} {\bibfnamefont {R.~M.}\ \bibnamefont {Ziff}}, \bibinfo {author} {\bibfnamefont {C.~D.}\ \bibnamefont {Lorenz}},\ and\ \bibinfo {author} {\bibfnamefont {P.}~\bibnamefont {Kleban}},\ }\bibfield  {title} {\bibinfo {title} {Shape-dependent universality in percolation},\ }\href {https://doi.org/https://doi.org/10.1016/S0378-4371(98)00569-X} {\bibfield  {journal} {\bibinfo  {journal} {Physica A}\ }\textbf {\bibinfo {volume} {266}},\ \bibinfo {pages} {17} (\bibinfo {year} {1999})}\BibitemShut {NoStop}%
\bibitem [{\citenamefont {Binder}(1981{\natexlab{a}})}]{PhysRevLett.47.693}%
  \BibitemOpen
  \bibfield  {author} {\bibinfo {author} {\bibfnamefont {K.}~\bibnamefont {Binder}},\ }\bibfield  {title} {\bibinfo {title} {Critical properties from {M}onte {C}arlo coarse graining and renormalization},\ }\href {https://doi.org/10.1103/PhysRevLett.47.693} {\bibfield  {journal} {\bibinfo  {journal} {Phys. Rev. Lett.}\ }\textbf {\bibinfo {volume} {47}},\ \bibinfo {pages} {693} (\bibinfo {year} {1981}{\natexlab{a}})}\BibitemShut {NoStop}%
\bibitem [{\citenamefont {Binder}(1981{\natexlab{b}})}]{binder1981finite}%
  \BibitemOpen
  \bibfield  {author} {\bibinfo {author} {\bibfnamefont {K.}~\bibnamefont {Binder}},\ }\bibfield  {title} {\bibinfo {title} {Finite size scaling analysis of ising model block distribution functions},\ }\href {https://doi.org/10.1007/BF01293604} {\bibfield  {journal} {\bibinfo  {journal} {Z. Phys. B}\ }\textbf {\bibinfo {volume} {43}},\ \bibinfo {pages} {119} (\bibinfo {year} {1981}{\natexlab{b}})}\BibitemShut {NoStop}%
\bibitem [{\citenamefont {Ouyang}\ \emph {et~al.}(2018)\citenamefont {Ouyang}, \citenamefont {Deng},\ and\ \citenamefont {Bl\"ote}}]{PhysRevE.98.062101}%
  \BibitemOpen
  \bibfield  {author} {\bibinfo {author} {\bibfnamefont {Y.}~\bibnamefont {Ouyang}}, \bibinfo {author} {\bibfnamefont {Y.}~\bibnamefont {Deng}},\ and\ \bibinfo {author} {\bibfnamefont {H.~W.~J.}\ \bibnamefont {Bl\"ote}},\ }\bibfield  {title} {\bibinfo {title} {Equivalent-neighbor percolation models in two dimensions: Crossover between mean-field and short-range behavior},\ }\href {https://doi.org/10.1103/PhysRevE.98.062101} {\bibfield  {journal} {\bibinfo  {journal} {Phys. Rev. E}\ }\textbf {\bibinfo {volume} {98}},\ \bibinfo {pages} {062101} (\bibinfo {year} {2018})}\BibitemShut {NoStop}%
\bibitem [{\citenamefont {Newman}\ and\ \citenamefont {Ziff}(2000)}]{PhysRevLett.85.4104}%
  \BibitemOpen
  \bibfield  {author} {\bibinfo {author} {\bibfnamefont {M.~E.~J.}\ \bibnamefont {Newman}}\ and\ \bibinfo {author} {\bibfnamefont {R.~M.}\ \bibnamefont {Ziff}},\ }\bibfield  {title} {\bibinfo {title} {Efficient {M}onte {C}arlo algorithm and high-precision results for percolation},\ }\href {https://doi.org/10.1103/PhysRevLett.85.4104} {\bibfield  {journal} {\bibinfo  {journal} {Phys. Rev. Lett.}\ }\textbf {\bibinfo {volume} {85}},\ \bibinfo {pages} {4104} (\bibinfo {year} {2000})}\BibitemShut {NoStop}%
\bibitem [{\citenamefont {Mendelson}(1999)}]{PhysRevE.60.6496}%
  \BibitemOpen
  \bibfield  {author} {\bibinfo {author} {\bibfnamefont {K.~S.}\ \bibnamefont {Mendelson}},\ }\bibfield  {title} {\bibinfo {title} {Percolation threshold of correlated two-dimensional lattices},\ }\href {https://doi.org/10.1103/PhysRevE.60.6496} {\bibfield  {journal} {\bibinfo  {journal} {Phys. Rev. E}\ }\textbf {\bibinfo {volume} {60}},\ \bibinfo {pages} {6496} (\bibinfo {year} {1999})}\BibitemShut {NoStop}%
\bibitem [{\citenamefont {Deng}\ \emph {et~al.}(2019)\citenamefont {Deng}, \citenamefont {Ouyang},\ and\ \citenamefont {Blöte}}]{deng2019medium}%
  \BibitemOpen
  \bibfield  {author} {\bibinfo {author} {\bibfnamefont {Y.}~\bibnamefont {Deng}}, \bibinfo {author} {\bibfnamefont {Y.}~\bibnamefont {Ouyang}},\ and\ \bibinfo {author} {\bibfnamefont {H.~W.~J.}\ \bibnamefont {Blöte}},\ }\bibfield  {title} {\bibinfo {title} {Medium-range percolation in two dimensions},\ }\href {https://doi.org/10.1088/1742-6596/1163/1/012001} {\bibfield  {journal} {\bibinfo  {journal} {J. Phys.: Conf. Ser.}\ }\textbf {\bibinfo {volume} {1163}},\ \bibinfo {pages} {012001} (\bibinfo {year} {2019})}\BibitemShut {NoStop}%
\bibitem [{\citenamefont {Feshanjerdi}\ and\ \citenamefont {Grassberger}(2024)}]{feshanjerdi2024}%
  \BibitemOpen
  \bibfield  {author} {\bibinfo {author} {\bibfnamefont {M.}~\bibnamefont {Feshanjerdi}}\ and\ \bibinfo {author} {\bibfnamefont {P.}~\bibnamefont {Grassberger}},\ }\bibfield  {title} {\bibinfo {title} {Extreme-value statistics and super-universality in critical percolation?},\ }\href {https://arxiv.org/abs/2401.05234} {\bibfield  {journal} {\bibinfo  {journal} {arXiv preprint arXiv:2401.05234}\ } (\bibinfo {year} {2024})}\BibitemShut {NoStop}%
\bibitem [{\citenamefont {Hu}\ \emph {et~al.}(2012)\citenamefont {Hu}, \citenamefont {Bl{\"o}te},\ and\ \citenamefont {Deng}}]{hu2012percolation}%
  \BibitemOpen
  \bibfield  {author} {\bibinfo {author} {\bibfnamefont {H.}~\bibnamefont {Hu}}, \bibinfo {author} {\bibfnamefont {H.~W.}\ \bibnamefont {Bl{\"o}te}},\ and\ \bibinfo {author} {\bibfnamefont {Y.}~\bibnamefont {Deng}},\ }\bibfield  {title} {\bibinfo {title} {Percolation in the canonical ensemble},\ }\href {https://doi.org/10.1088/1751-8113/45/49/494006} {\bibfield  {journal} {\bibinfo  {journal} {J. Phys. A: Math. Theor.}\ }\textbf {\bibinfo {volume} {45}},\ \bibinfo {pages} {494006} (\bibinfo {year} {2012})}\BibitemShut {NoStop}%
\bibitem [{\citenamefont {Qian}\ \emph {et~al.}(2005)\citenamefont {Qian}, \citenamefont {Deng},\ and\ \citenamefont {Bl\"ote}}]{PhysRevB.71.144303}%
  \BibitemOpen
  \bibfield  {author} {\bibinfo {author} {\bibfnamefont {X.}~\bibnamefont {Qian}}, \bibinfo {author} {\bibfnamefont {Y.}~\bibnamefont {Deng}},\ and\ \bibinfo {author} {\bibfnamefont {H.~W.~J.}\ \bibnamefont {Bl\"ote}},\ }\bibfield  {title} {\bibinfo {title} {Percolation in one of $q$ colors near criticality},\ }\href {https://doi.org/10.1103/PhysRevB.71.144303} {\bibfield  {journal} {\bibinfo  {journal} {Phys. Rev. B}\ }\textbf {\bibinfo {volume} {71}},\ \bibinfo {pages} {144303} (\bibinfo {year} {2005})}\BibitemShut {NoStop}%
\bibitem [{\citenamefont {Deng}\ and\ \citenamefont {Bl\"ote}(2005)}]{PhysRevE.71.016117}%
  \BibitemOpen
  \bibfield  {author} {\bibinfo {author} {\bibfnamefont {Y.}~\bibnamefont {Deng}}\ and\ \bibinfo {author} {\bibfnamefont {H.~W.~J.}\ \bibnamefont {Bl\"ote}},\ }\bibfield  {title} {\bibinfo {title} {Surface critical phenomena in three-dimensional percolation},\ }\href {https://doi.org/10.1103/PhysRevE.71.016117} {\bibfield  {journal} {\bibinfo  {journal} {Phys. Rev. E}\ }\textbf {\bibinfo {volume} {71}},\ \bibinfo {pages} {016117} (\bibinfo {year} {2005})}\BibitemShut {NoStop}%
\bibitem [{\citenamefont {Lu}\ \emph {et~al.}(2024{\natexlab{b}})\citenamefont {Lu}, \citenamefont {Song}, \citenamefont {Li},\ and\ \citenamefont {Deng}}]{lu2024}%
  \BibitemOpen
  \bibfield  {author} {\bibinfo {author} {\bibfnamefont {M.}~\bibnamefont {Lu}}, \bibinfo {author} {\bibfnamefont {Y.-F.}\ \bibnamefont {Song}}, \bibinfo {author} {\bibfnamefont {M.}~\bibnamefont {Li}},\ and\ \bibinfo {author} {\bibfnamefont {Y.}~\bibnamefont {Deng}},\ }\bibfield  {title} {\bibinfo {title} {Self-similar gap dynamics in percolation and rigidity percolation},\ }\href {https://arxiv.org/abs/2411.04748} {\bibfield  {journal} {\bibinfo  {journal} {arXiv preprint arXiv:2411.04748}\ } (\bibinfo {year} {2024}{\natexlab{b}})}\BibitemShut {NoStop}%
\bibitem [{\citenamefont {Beirlant}\ \emph {et~al.}(2006)\citenamefont {Beirlant}, \citenamefont {Goegebeur}, \citenamefont {Segers},\ and\ \citenamefont {Teugels}}]{beirlant2006statistics}%
  \BibitemOpen
  \bibfield  {author} {\bibinfo {author} {\bibfnamefont {J.}~\bibnamefont {Beirlant}}, \bibinfo {author} {\bibfnamefont {Y.}~\bibnamefont {Goegebeur}}, \bibinfo {author} {\bibfnamefont {J.}~\bibnamefont {Segers}},\ and\ \bibinfo {author} {\bibfnamefont {J.~L.}\ \bibnamefont {Teugels}},\ }\href@noop {} {\emph {\bibinfo {title} {Statistics of extremes: Theory and applications}}}\ (\bibinfo  {publisher} {John Wiley \& Sons},\ \bibinfo {year} {2006})\BibitemShut {NoStop}%
\end{thebibliography}%

\end{document}